\documentclass[12pt]{article}
\makeatletter
\usepackage[dvips]{graphics}
\newcommand{\singlespacing}{\let\CS=\@currsize\renewcommand{\baselinestretch}
{1.0}\tiny\CS}
\newcommand{\doublespacing}{\let\CS=\@currsize\renewcommand{\baselinestretch}
{1.5}\tiny\CS}

\begin{document}

\title{Nonlocality without inequality for almost all two-qubit entangled state
 based on Cabello's nonlocality argument}
\author{Samir Kunkri\thanks{{\bf E-mail:skunkri$\_$r@isical.ac.in}}
\thinspace , Sujit K. Choudhary \thanks{{\bf
E-mail:sujit$\_$r@isical.ac.in}}\\
Physics and Applied Mathematics Unit, \\
Indian Statistical Institute, \\
Kolkata 700108 , India\\ and\\ Ali Ahanj and Promod Joag
\thanks{{\bf E-mail:pramod@physics.unipune.ernet.in}}\\
Department of Physics, University of Pune,\\ Pune 411007, India }

\maketitle

\vspace{0.5cm}
%\noindent
\begin{center}
{\bf Abstract}
\end{center}

{\small Here we deal with a nonlocality argument proposed by
Cabello which is more general than Hardy's nonlocality argument
but still maximally entangled states do not respond. However, for
most of the other entangled states maximum probability of success
of this argument is more than that of the Hardy's argument.}\\

\section*{Introduction}

It is well known result that realistic interpretations of quantum
theory are nonlocal \cite{bell}. This was first shown by means of
Bell's inequality. Afterwards, the proof of the same for three
spin-1/2 particles as well as for two spin-1 particles, without
using inequality caused much interest among physicists
\cite{green}. Surprisingly Hardy gave a proof of nonlocality
without using inequality, for two spin-1/2 particles which
requires two measurement settings on  both the sides as happens in
case of Bell's argument \cite{hardy92}. Later  Hardy showed this
kind of nonlocality argument can be made for almost all entangled
state of two spin-1/2 particles except for maximally entangled
one.\cite{hardy93}. He considered the cases where the measurement
choices were  same for both the parties. Jordan showed that for
any given entangled state of two spin-1/2 particles except
maximally entangled state there are  many set  of observables on
each side which satisfy Hardy's nonlocality conditions
\cite{jordan}. Jordan also showed that the set of observables
which gives maximum probability of success in showing the
contradiction with local-realism, is the same as
chosen by Hardy.\\
Recently Cabello has introduced a logical structure to prove
Bell's theorem without inequality for three particles GHZ and W
state \cite{cabello02}. Logical structure presented by  Cabello is
as follows : Consider four events D, E, F and G where D and F may
happen in one system and E and G happen in another system which is
far apart from the first. The probability of joint occurrence of D
and E is non-zero, E always implies F, D always implies G, but F
and G happen with lower probability than D and E. These four
statements are not compatible with local realism. The difference
between these two probabilities is the measure of violation of
local realism. Though Cabello's logical structure was originally
proposed for showing nonlocality for three particle states but
Liang and Li \cite{liang05} exploited it in establishing
nonlocality without inequality for a class of two qubit mixed
entangled state. In this sense, Hardy's logical structure is an
special case of Cabello's structure as the logical structure of
Hardy for establishing nonlocality is as follows: D and E
sometimes happen, E always implies F, D always implies G, but F
and G never happen. Recently based on Cabello's logical structure
Kunkri and Choudhary \cite{kunkri} have shown that there may be
many classes of two qubit mixed states which exhibit nonlocality
without inequality. It is noteworthy here that in contrast there
is no two qubit mixed state which shows Hardy type nonlocality
\cite{karpra}. So it seems interesting to study that whether
maximally entangled states follow this more general (than
Hardy's), Cabello's nonlocality argument or not , because Hardy's
nonlocality argument is not followed by a maximally entangled
state. In this paper we have studied it and found that maximally
entangled states do not respond even to this argument. However,
for all other pure entangled states , Cabello's argument runs. We
further have enquired about the highest value of difference
between the two probabilities which appear in Cabello's argument.
Surprisingly this value differs from the highest value of
probability which
appears in Hardy's argument.\\
\section*{Cabello's argument for two qubits}
Let us consider two spin-1/2 particles A and B. Let F, D, G and E
represent the spin observables along $n_F (
\sin{\theta_F}\cos{\phi_F}, \sin{\theta_F}\sin{\phi_F},
\cos{\theta_F})$, $n_D ( \sin{\theta_D}\cos{\phi_D},
\sin{\theta_D}\sin{\phi_D}, \cos{\theta_D})$, $n_G (
\sin{\theta_G}\cos{\phi_G}, \sin{\theta_G}\sin{\phi_G},
\cos{\theta_G})$ and $n_E ( \sin{\theta_E}\cos{\phi_E},
\sin{\theta_E}\sin{\phi_E}, \cos{\theta_E})$ respectively. Every
observable has the eigen value $\pm 1$. Let F and D are measured
on particle A and G and E are measured on particle B. Now we
consider the following equations
\begin{equation}
P(F = +1, G = +1) = q_1
\end{equation}
\begin{equation}
P(D = +1, G = -1) = 0
\end{equation}
\begin{equation}
P(F = -1, E = +1) = 0
\end{equation}
\begin{equation}
P(D = +1, E = +1) = q_4
\end{equation}

Equation (1) tells that if F is measured on particle A and G is
measured on particle B, then the probability that both will get +1
eigen value is $q_1$. Other equations can be analyzed in a similar
fashion. These equations form the basis of Cabello's nonlocality
argument. It can easily be seen that these equations contradict
local-realism if $q_1 < q_4$. To show this, let us consider those
hidden variable states $\lambda$ for which $D = +1$ and $E = +1$.
Now for these states  equations $(2)$ and $(3)$ tell that the
values of $G$ and $F$ must be equal to $+1$. Thus according to
local realism $P(F = +1, G = +1)$ should be at least equal to
$q_4$, which contradicts equation $(1)$ as $q_1 < q_4$. It should
be noted here that $q_1=0$ reduces this argument to Hardy's one.
So by Cabello's argument we specifically mean that the above
argument runs even with nonzero $q_1$.\\
Now we will show that for almost all two qubit pure entangled
state other than maximally entangled one this kind of nonlocality
 argument runs. Following Schmidt
 decomposition procedure  any
 entangled state of two particles A and B can be written as
 \begin{equation}
 |\psi\rangle = (\cos{\beta}) |0\rangle_A |0\rangle_B + (\sin
 {\beta})e^{i\gamma} |1\rangle_A |1\rangle_B
 \end{equation}
If either $\cos{\beta}$ or $\sin{\beta}$ is zero, we have a
product state not an entangled state. Then it is not possible to
satisfy equation $(1)-(4)$. Hence we assume that neither
$\cos{\beta}$ nor $\sin{\beta}$ is zero; both are positive.\\
The density matrix for the above state is
\begin{equation}
\begin{array}{lcl}
 \rho = \frac{1}{4}[I^A \otimes I^B + (\cos^2{\beta} - \sin^2{\beta})I^A\otimes \sigma_z^B +
(\cos^2{\beta} - \sin^2{\beta})\sigma_z^A \otimes I^B \\ +
(2\cos{\beta}\sin{\beta}\cos{\gamma})\sigma_x^A \otimes \sigma_x^B
+ (2\cos{\beta}\sin{\beta}\sin{\gamma})\sigma_x^A \otimes
\sigma_y^B \\ + (2\cos{\beta}\sin{\beta}\sin{\gamma})\sigma_y^A
\otimes \sigma_x^B -
(2\cos{\beta}\sin{\beta}\cos{\gamma})\sigma_y^A \otimes \sigma_y^B
+ \sigma_z^A \otimes \sigma_z^B]
\end{array}
\end{equation}

Where $\sigma_x$, $\sigma_y$ and $\sigma_z$ are Pauli operators.
Now for this state if F is measured on particle A and G is
measured on particle B, then the probability that both will get +1
eigen value is given by

\begin{equation}
\begin{array}{lcl}
 P(F = +1, G = +1) = (\frac{1}{4})[1 + (\cos^2{\beta} -
 \sin^2{\beta})(\cos{\theta_F} + \cos{\theta_G})\\
  + \cos{\theta_F}\cos{\theta_G} + 2\cos{\beta}\sin{\beta}\sin{\theta_F}\sin{\theta_G}
  \times \cos{(\phi_F + \phi_G - \gamma)}]
\end{array}
\end{equation}
Rearranging the above expression we get
\begin{equation}
\begin{array}{lcl}
 P(F = +1, G = +1) =
 \cos^2{\beta}\cos^2{\frac{\theta_F}{2}}\cos^2{\frac{\theta_G}{2}} +
 \sin^2{\beta}\sin^2{\frac{\theta_F}{2}}\sin^2{\frac{\theta_G}{2}} +
 \\
  + 2\cos{\beta}\sin{\beta}\cos{\frac{\theta_F}{2}}\sin{\frac{\theta_F}{2}}
  \cos{\frac{\theta_G}{2}}\sin{\frac{\theta_G}{2}}
  \times \cos{(\phi_F + \phi_G - \gamma)}]= q_1(say)
\end{array}
\end{equation}
Similar calculations for other probabilities give us:
\begin{equation}
\begin{array}{lcl}
 P(D = +1, G = -1) =
 \cos^2{\beta}\cos^2{\frac{\theta_D}{2}}\sin^2{\frac{\theta_G}{2}} +
 \sin^2{\beta}\sin^2{\frac{\theta_D}{2}}\cos^2{\frac{\theta_G}{2}} +
 \\
  + 2\cos{\beta}\sin{\beta}\cos{\frac{\theta_D}{2}}\sin{\frac{\theta_D}{2}}
  \cos{\frac{\theta_G}{2}}\sin{\frac{\theta_G}{2}}
  \times \cos{(\phi_D + \phi_G + \pi - \gamma)}]= q_2(say)
\end{array}
\end{equation}

\begin{equation}
\begin{array}{lcl}
 P(F = -1, E = +1) =
 \cos^2{\beta}\cos^2{\frac{\theta_E}{2}}\sin^2{\frac{\theta_F}{2}} +
 \sin^2{\beta}\sin^2{\frac{\theta_E}{2}}\cos^2{\frac{\theta_F}{2}} +
 \\
  + 2\cos{\beta}\sin{\beta}\cos{\frac{\theta_F}{2}}\sin{\frac{\theta_F}{2}}
  \cos{\frac{\theta_E}{2}}\sin{\frac{\theta_E}{2}}
  \times \cos{(\phi_F + \phi_E + \pi - \gamma)}]=q_3(say)
\end{array}
\end{equation}

\begin{equation}
\begin{array}{lcl}
 P(D = +1, E = +1) =
 \cos^2{\beta}\cos^2{\frac{\theta_D}{2}}\cos^2{\frac{\theta_E}{2}} +
 \sin^2{\beta}\sin^2{\frac{\theta_D}{2}}\sin^2{\frac{\theta_E}{2}} +
 \\
  + 2\cos{\beta}\sin{\beta}\cos{\frac{\theta_D}{2}}\sin{\frac{\theta_D}{2}}
  \cos{\frac{\theta_E}{2}}\sin{\frac{\theta_E}{2}}
  \times \cos{(\phi_D + \phi_E - \gamma)}]=q_4(say)
\end{array}
\end{equation}
For running  Cabello's nonlocality argument, following conditions
should be satisfied:
\begin{equation}
q_2=0,~~ q_3=0,~~ (q_4 - q_1) > 0, ~~ q_1 > 0
\end{equation}
%Now $q_2$ will be equal to zero if any one of the following
%conditions is satisfied:\\
%(a) $cos{\frac{\theta_D}{2}}=0 $ and $ cos{\frac{\theta_G}{2}}=0$\\
%(b) $sin{\frac{\theta_D}{2}}=0 $ and $ sin{\frac{\theta_G}{2}}=0$\\
%Similarly $q_1$ will be equal to zero if any one of the following
%conditions is satisfied:\\
%(a) $cos{\frac{\theta_E}{2}}=0 $ and $ cos{\frac{\theta_F}{2}}=0$\\
%(b) $sin{\frac{\theta_E}{2}}=0 $ and $ sin{\frac{\theta_F}{2}}=0$\\
 %So in all there will be four combinations to make both $q_2$ and
%$q_1$
%equal to zero namely:\\
%(a) $cos{\frac{\theta_D}{2}}=0 $ and $ cos{\frac{\theta_G}{2}}=0$\\
 %   $cos{\frac{\theta_E}{2}}=0 $ and $ cos{\frac{\theta_F}{2}}=0$\\
%(b) $cos{\frac{\theta_D}{2}}=0 $ and $ cos{\frac{\theta_G}{2}}=0$\\
 %   $sin{\frac{\theta_E}{2}}=0 $ and $ sin{\frac{\,}{2}}=0$\\
%(c) $sin{\frac{\theta_D}{2}}=0 $ and $ sin{\frac{\theta_G}{2}}=0$\\
 %   $sin{\frac{\theta_E}{2}}=0 $ and $ sin{\frac{\theta_F}{2}}=0$\\
%(d) $sin{\frac{\theta_D}{2}}=0 $ and $ sin{\frac{\theta_G}{2}}=0$\\
 %   $cos{\frac{\theta_E}{2}}=0 $ and $ cos{\frac{\theta_F}{2}}=0$\\
 %One can easily check by putting any of the above combinations in (12)that
 %no combination can give $$(q_4 - q_1) > 0$$
%In fact to fulfill the conditions given by (12),neither of the
%$\cos{\frac{\theta}{2}}$ and  $\sin{\frac{\theta}{2}}$ should be
%zero for any of the angles
%$\theta_D$,$\theta_E$,$\theta_G$,$\theta_F$. We assume further
%that every $\cos{\frac{\theta}{2}}$ and  $\sin{\frac{\theta}{2}}$
%is positive. \\
Since $q_2$ represents probability, it can not be negative. If it
is zero, it is at its minimum value. Then its derivative must be
zero. From it's derivative with respect to $\phi_D$ we see that
$\sin{(\phi_D + \phi_G + \pi - \gamma)}$ must be zero. Evidently

\begin{equation}
\cos{(\phi_D + \phi_G + \pi - \gamma)} = -1
\end{equation}
We conclude that if  $q_2$ is zero, then
\begin{equation}
\cos{\beta}\cos{\frac{\theta_D}{2}}\sin{\frac{\theta_G}{2}} =
\sin{\beta}\sin{\frac{\theta_D}{2}}\cos{\frac{\theta_G}{2}}
\end{equation}
Similar sort of argument for $q_3$ to be zero will give:
\begin{equation}
\cos{(\phi_F + \phi_E + \pi - \gamma)} = -1
\end{equation}
and
\begin{equation}
\cos{\beta}\cos{\frac{\theta_E}{2}}\sin{\frac{\theta_F}{2}} =
\sin{\beta}\sin{\frac{\theta_E}{2}}\cos{\frac{\theta_F}{2}}
\end{equation}
\section*{Maximally entangled states of two spin-1/2 particles do not
exhibit Cabello type nonlocality-}
 For maximally entangled state
$\tan{\beta} = 1$, then from equations $(14)$ and $(16)$ we get
\begin{equation}
\frac{\theta_G}{2} = \frac{\theta_D}{2} + n\pi
\end{equation}
\begin{equation}
\frac{\theta_F}{2} = \frac{\theta_E}{2} + m\pi
\end{equation}
Using equations $(17)$ and $(18)$ first in equation $(8)$ and then
in equation (11) we get $q_1$ and $q_4$ for maximally entangled
state as:
\begin{equation}
\begin{array}{lcl}
q_1 =
 \frac{1}{2}\cos^2{\frac{\theta_D}{2}}\cos^2{\frac{\theta_E}{2}} +
 \frac{1}{2}\sin^2{\frac{\theta_D}{2}}\sin^2{\frac{\theta_E}{2}}
 \\
  + \cos{\frac{\theta_D}{2}}\sin{\frac{\theta_D}{2}}
  \cos{\frac{\theta_E}{2}}\sin{\frac{\theta_E}{2}}
  \times \cos{(\phi_F + \phi_G - \gamma)}]
\end{array}
\end{equation}

\begin{equation}
\begin{array}{lcl}
q_4 =
 \frac{1}{2}\cos^2{\frac{\theta_D}{2}}\cos^2{\frac{\theta_E}{2}} +
 \frac{1}{2}\sin^2{\frac{\theta_D}{2}}\sin^2{\frac{\theta_E}{2}}
 \\
  + \cos{\frac{\theta_D}{2}}\sin{\frac{\theta_D}{2}}
  \cos{\frac{\theta_E}{2}}\sin{\frac{\theta_E}{2}}
  \times \cos{(\phi_D + \phi_E - \gamma)}]
\end{array}
\end{equation}
From equations $(19)$ and $(20)$ it is clear that $q_4$ will be
grater than $ q_1$ for a maximally entangled state  only when
$\cos{(\phi_D + \phi_E - \gamma)}
> \cos{(\phi_F + \phi_G - \gamma)}$. But equation $(13)$ together
with equation $(15)$ says that $\cos{(\phi_D + \phi_E - \gamma)} =
\cos{(\phi_F + \phi_G - \gamma)}$ {\it i.e} $ q_4 = q_1$. So one
can conclude that there is no choice of observable which can make
maximally entangled state to show Cabello type of
nonlocality .\\
\section*{Cabello's argument runs for other two particle pure
entangled states-}
 To show  that for every pure entangled state other than maximally
 entangled state of two spin-1/2 particles, Cabello like argument runs
 it will be sufficient to show that one can always choose a set of observables for which
 set of conditions given
by equation (12) is satisfied. This is equivalent of saying that
for $ 0<\beta<\frac{\pi}{2}$ except when $\beta=\frac{\pi}{4}$
there is at least one value for each of
$\theta_D$,$\theta_E$,$\theta_G$,$\theta_F,\phi_D$,$\phi_E$,$\phi_G$,$\phi_F$
for which  conditions mentioned in(12) are satisfied.\\
Let us choose our $\phi's$ in such a manner that\\
$$cos{(\phi_F + \phi_G - \gamma)}= cos{(\phi_D + \phi_E -
\gamma)}=-1$$
For these $\phi's$ equations (8) and (11) respectively will read
as:\\
\begin{equation}
q_1 = (\cos{\beta}\cos{\frac{\theta_F}{2}}\cos{\frac{\theta_G}{2}}
- \sin{\beta}\sin{\frac{\theta_F}{2}}\sin{\frac{\theta_G}{2}})^2
\end{equation}
\begin{equation}
q_4 = (\cos{\beta}\cos{\frac{\theta_D}{2}}\cos{\frac{\theta_E}{2}}
- \sin{\beta}\sin{\frac{\theta_D}{2}}\sin{\frac{\theta_E}{2}})^2
\end{equation}
So
\begin{equation}
\begin{array}{lcl}
 (q_4 - q_1) =
 \cos^2{\beta}(\cos^2{\frac{\theta_D}{2}}\cos^2{\frac{\theta_E}{2}}
 - \cos^2{\frac{\theta_F}{2}}\cos^2{\frac{\theta_G}{2}}) +
 \sin^2{\beta}(\sin^2{\frac{\theta_D}{2}}\sin^2{\frac{\theta_E}{2}}
 - \sin^2{\frac{\theta_F}{2}}\sin^2{\frac{\theta_G}{2}})\\
 + 2\sin{\beta}\cos{\beta}(\cos{\frac{\theta_F}{2}}\cos{\frac{\theta_G}{2}}\sin{\frac{\theta_F}{2}}
 \sin{\frac{\theta_G}{2}} - \cos{\frac{\theta_D}{2}}\cos{\frac{\theta_E}{2}}\sin{\frac{\theta_D}{2}}
 \sin{\frac{\theta_E}{2}})
\end{array}
\end{equation}
Now we will have to choose at least one set of values of
$\theta's$ in such a way that $(q_4 - q_1)$ and $q_1$ are nonzero
and positive. Moreover, these values of $\theta's$ should also not
violate conditions given in equations $(14)$ and $(16)$.\\
%For convenience,we consider the following two cases separately:\\
%(a)$\tan{\beta} > 1$  and  (b)  $\tan{\beta} < 1$
%\\For $\tan{\beta}> 1$,
let us try with $ \frac{\theta_D}{2} = 0$ {\it i.e}
$$ \sin{\frac{\theta_D}{2}} = 0,~~~\cos{\frac{\theta_D}{2}} = 1$$
This makes equation $(14)$ to read as
 $$\sin{\frac{\theta_G}{2}} = 0,\Rightarrow {\frac{\theta_G}{2}} =
 0$$
 Then from equation $(23)$ we get $$ (q_4 - q_1) =
 \cos^2{\beta}(\cos^2{\frac{\theta_E}{2}}-
 \cos^2{\frac{\theta_F}{2}})$$
Thus $(q_4 - q_1) > 0$ if
\begin{equation}
\cos{\frac{\theta_E}{2}}> \cos{\frac{\theta_F}{2}}
\end{equation}
 Rewriting equation $(16)$ as
 \begin{equation}
 \tan{\frac{\theta_F}{2}} = \tan{\beta}\tan{\frac{\theta_E}{2}}
\end{equation}
 Values of $\theta's$ satisfying inequality (24) will not violate
 equation (25) provided $\tan{\beta} > 1$.
Now for these values of $\theta's$, from equation (21), we get:
$q_1 = (\cos{\beta}\cos{\frac{\theta_F}{2}})^2$
which is greater than zero.\\
 So for the above values of $\theta's$ {\it i.e} for
$\frac{\theta_D}{2} = \frac{\theta_G}{2} = 0$ and
$\cos{\frac{\theta_E}{2}}> \cos{\frac{\theta_F}{2}}$, all the
states for which $\tan{\beta} > 1$ ; Cabello's nonlocality
argument runs.\\
For  other states {\it i.e} for the states for which $\tan{\beta}
< 1$, let us choose $\frac{\theta_D}{2} = \frac{\theta_G}{2} =
\frac {\pi}{2}$. Then from equation $(23)$ we get $$ (q_4 - q_1) =
 \sin^2{\beta}(\sin^2{\frac{\theta_E}{2}}-
 \sin^2{\frac{\theta_F}{2}})$$
 Thus $(q_4 - q_1) > 0$ if
\begin{equation}
\sin{\frac{\theta_E}{2}}> \sin{\frac{\theta_F}{2}}
\end{equation}
One can easily check that for abovementioned values of $\theta's$
; $q_1$ is also positive and equation (25) is satisfied too.

%Under this condition we have, $\tan{\frac{\theta_F}{2}} <
%\tan{\frac{\theta_E}{2}}$. Then to satisfy equation $(25)$,
%$\tan{\beta}$ must be less than $1$. {\it i.e } $\tan{\beta} < 1$.
Thus if we choose $\frac{\theta_D}{2} = \frac{\theta_G}{2} =
\frac{\pi}{2}$ and $\sin{\frac{\theta_E}{2}}>
\sin{\frac{\theta_F}{2}}$, then all the states for which,
$\tan{\beta} < 1$ satisfy Cabello's nonlocality argument. So for
every $\beta$ (except for $\beta=\frac{\pi}{4}$); we can choose
$\theta's$ and $\phi's$ and hence the observables in such a way
that Cabello's argument
runs.\\
\section*{Maximum probability of success}
For getting maximum probability of success of Cabello's argument
in contradicting local-realism we will have to maximize the
quantity $(q_4 - q_1)$ for a given $\beta$ over all observable
parameters $\theta's$ and $\phi's$ under the restrictions given by
equation's $(13)-(16)$. Using the equations $(13)-(16)$, we have
\begin{equation}
\begin{array}{lcl}
 (q_4 - q_1) =
 \cos^2{\beta}[(k_2 - k_1) + \tan^2{\beta}\tan^2{\frac{\theta_D}{2}}
 \tan^2{\frac{\theta_E}{2}} (k_2 - k_1 \tan^4{\beta}) + \\ 2 \tan{\beta}\tan{\frac{\theta_D}{2}}
 \tan{\frac{\theta_E}{2}} (k_2 - k_1 \tan^2{\beta})\cos{(\phi_D + \phi_E -
 \gamma)}]
\end{array}
\end{equation}
where $$ k_1 = \frac{1}
 {(\tan^2{\beta}\tan^2{\frac{\theta_D}{2}} + 1)(\tan^2{\beta}\tan^2{\frac{\theta_E}{2}} +
 1)},~~~~~ k_2 = \frac{1}
 {(\tan^2{\frac{\theta_D}{2}} + 1)(\tan^2{\frac{\theta_E}{2}} +
 1)} $$
 It is clear from the equation $(27)$ that one can obtain maximum value of  $(q_4 - q_1)$,
 when $\cos{(\phi_D + \phi_E - \gamma)}= \pm 1$.
Let us first consider $\cos{(\phi_D + \phi_E - \gamma)}= -1$, then
from equation $(27)$ we have
\begin{equation}
\begin{array}{lcl}
 (q_4 - q_1) =
 \cos^2{\beta}[\frac{(1- \tan{\beta}\tan{\frac{\theta_D}{2}}\tan{\frac{\theta_E}{2}})^2}
 {(\tan^2{\frac{\theta_D}{2}} + 1)(\tan^2{\frac{\theta_E}{2}} +
 1)} -
\frac{(1-
\tan^3{\beta}\tan{\frac{\theta_D}{2}}\tan{\frac{\theta_E}{2}})^2}
 {(\tan^2{\beta}\tan^2{\frac{\theta_D}{2}} + 1)(\tan^2{\beta}\tan^2{\frac{\theta_E}{2}} +
 1)} ]
\end{array}
\end{equation}
From the above equation one can show that $(q_4 - q_1)$ will be
maximum when $\theta_D = \theta_E$ (see Appendix) which in turn
implies  $\theta_G = \theta_F$ {\it i.e } $(q_4 - q_1)$ becomes
maximum when measurement settings in both the sides is  same as
was in Hardy's case. Now for the optimal case {\it i.e } for
$\theta_G = \theta_F$  and $\theta_D = \theta_E$, $(q_4 - q_1)$
becomes
\begin{equation}
\begin{array}{lcl}
 (q_4 - q_1) =
 \cos^2{\beta}[\frac{(1- \tan{\beta}\tan^2{\frac{\theta_D}{2}})^2}
 {(\tan^2{\frac{\theta_D}{2}} + 1)^2} -
\frac{(1- \tan^3{\beta}\tan^2{\frac{\theta_D}{2}})^2}
 {(\tan^2{\beta}\tan^2{\frac{\theta_D}{2}} + 1)^2} ]
\end{array}
\end{equation}
Numerically we have checked that $(q_4 - q_1)$ has a maximum value
of $.1078$ when $\cos{\beta} = .485$ with $\theta_D = \theta_E =
.59987$. This is interesting as maximum probability of success of
 Hardy's argument is only $9\%$, whereas in case of Cabello's
argument it is approximately $11\%$.\\ Here we are  comparing the
maximum probability of success  of Hardy's argument with that of
Cabello's argument for all states.\\
\begin{figure}[hp]
\begin{center}
\scalebox{0.6}{\includegraphics{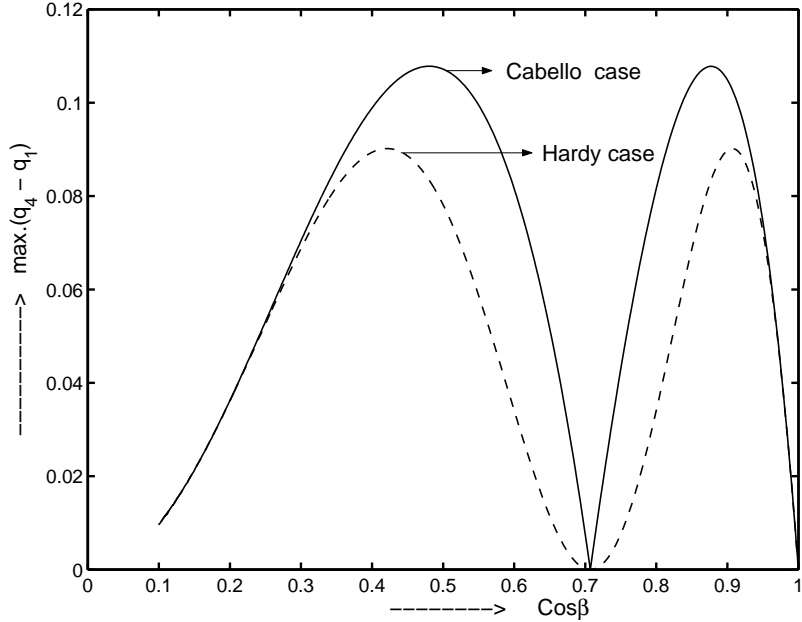}}
\end{center}
\caption{Comparison of the maximum probability of success between
Hardy's and Cabello's case}
\end{figure}

Graph shows that for $\cos{\beta} \approx.7$ {\it i.e } for
$\beta=\frac{\pi}{4}$ and for $\cos{\beta}=1$ {\it i.e } for
$\beta=0$; maximum of  $(q_4 - q_1)$ vanishes. This is as expected
because these values of $\beta$ represent respectively the
maximally entangled and product states for which Cabello's
argument does not run. For most of the other values of $\beta$
{\it i.e } for most of the other entangled states , maximum
 probability of success of Cabello's argument in establishing their
 nonlocal feature is more than the maximum probability of success
 of hardy's argument in doing the same.\\
As we have mentioned earlier (just before equation 28) that
$\cos{(\phi_D + \phi_E - \gamma)}= 1$ also optimizes $(q_4 -
q_1)$. This also gives the same maximum value for $(q_4 - q_1)$as
given by $\cos{(\phi_D + \phi_E - \gamma)}= -1$
but for $\theta_D = -\theta_E$.\\
\section*{Conclusion}

In conclusion, here we have shown that maximally entangled states
do not respond even to Cabello's argument which is a relaxed one
and is more general than Hardy's argument. All other pure
entangled states response to Cabello's argument. These states also
exhibit Hardy type nonlocality. But, interestingly for most of
these nonmaximally entangled states, fraction of runs in which
Cabello's argument succeeds in demonstrating their nonlocal
feature can be made more than the fraction of runs in which
Hardy's argument succeeds in doing the same. So it seems that in
some sense, for demonstrating the  nonlocal features of  most of
the entangled
states, Cabello's argument is a better candidate.\\
{\bf Appendix-}\\ We want to optimize $(q_4 - q_1)$ given in
equation $(28)$ with respect to $\theta_D$ and $\theta_E$ for a
given $\beta$. Differentiating equation $(28)$ with respect to
$\theta_D$ and equating it to zero, we have the following two
equations
\begin{equation}
(\tan{\beta}\tan{\frac{\theta_E}{2}} + \tan {\frac{\theta_D}{2}})=
0
\end{equation}
and
\begin{equation}
(\tan{\beta}\tan{\frac{\theta_E}{2}}\tan{\frac{\theta_D}{2}} -
1)(\tan^2{\beta}\tan^2{\frac{\theta_E}{2}} +
1)(\tan^2{\beta}\tan^2{\frac{\theta_D}{2}} + 1)^2 = $$
$$ (\tan^2{\beta}\sec^2{\frac{\theta_D}{2}})(\tan^3{\beta}\tan{\frac{\theta_E}{2}}\tan{\frac{\theta_D}{2}}
- 1)(\sec^2{\frac{\theta_E}{2}}\sec^2{\frac{\theta_D}{2}})
\end{equation}
 Similarly differentiating equation $(28)$ with
respect to $\theta_E$ and equating it to zero, we have
\begin{equation}
(\tan{\beta}\tan{\frac{\theta_D}{2}} + \tan {\frac{\theta_E}{2}})=
0
\end{equation}
and
\begin{equation}
(\tan{\beta}\tan{\frac{\theta_D}{2}}\tan{\frac{\theta_E}{2}} -
1)(\tan^2{\beta}\tan^2{\frac{\theta_D}{2}} +
1)(\tan^2{\beta}\tan^2{\frac{\theta_E}{2}} + 1)^2 = $$
$$ (\tan^2{\beta}\sec^2{\frac{\theta_E}{2}})(\tan^3{\beta}\tan{\frac{\theta_D}{2}}\tan{\frac{\theta_E}{2}}
- 1)(\sec^2{\frac{\theta_D}{2}}\sec^2{\frac{\theta_E}{2}})
\end{equation}
Analyzing above four conditions we have
$$\theta_D = \theta_E$$ will give the optimal solution.
Similarly for $cos{(\phi_D + \phi_E - \gamma)}= +1$, we will get
same kind of results.
\section*{Acknowledgement}
Authors would like to thank Guruprasad Kar, Debasis Sarkar for
useful discussions. We also thank Swarup Poria to help us in
numerical calculation. S.K acknowledges the support by the Council
of Scientific and Industrial Research, Government of India, New
Delhi.


\begin{thebibliography}{99}
\bibitem{bell} J.S. Bell, {\it Physics} {\bf 1}, 195 (1964).
\bibitem{green} D.M. Greenberger, M.A. Horne and A. Zeilinger, in:
{\it Bell's theorem, quantum theory and conceptions of the
universe}, edited by M. kafatos (Kluwer, Dordrecht, 1989) p. 69.
\bibitem{hardy92} L. Hardy, {\it Phys. Rev. Lett. } {\bf 68}, 2981 (1992).
\bibitem{hardy93} L. Hardy, {\it Phys. Rev. Lett. } {\bf 71}, 1665 (1993).
\bibitem{jordan} T. F. Jordan, {\it Phys. Rev. A} {\bf 50}, 62 (1994).
\bibitem{cabello02} A. Cabello,  {\it Phys. Rev. A} {\bf 65}, 032108 (2002).
\bibitem{liang05} Lin-mei Liang and Cheng-zu Li, {\it Phys. Lett. A} {\bf 335 }, 371 (2005).
\bibitem{kunkri} S. Kunkri and S.K. Choudhary {\it Phys. Rev. A} {\bf 72}, 022348 (2005).
\bibitem{karpra} G. Kar, {\it Phys. Rev. A} {\bf 56}, 1023 (1997).
\end{thebibliography}
\end{document}